\documentclass[3p,preprint]{elsarticle}
\usepackage{verbatim} 
\usepackage{dsfont}
\usepackage{amsmath,amssymb}
\usepackage{epsfig}
\usepackage[active]{srcltx}
\usepackage{wrapfig}
\usepackage{cancel}

\def\x'{\mathaccent 19 x}
\def\y'{\mathaccent 19 y}
\def\n'{\mathaccent 19 n}
\def\u'{\mathaccent 19 u}
\def\et'{\mathaccent 19 \eta}
\def\th'{\mathaccent 19 \theta}
\def\lam'{\mathaccent 19 \lambda}
\def\varet'{\mathaccent 19 \vartheta}
\def\rh'{\mathaccent 19 \rho}
\def\ph'{\mathaccent 19 \Phi}
\def\xb'{\mathaccent 19 {\bar{x}}}

\def\bsp{\be\begin{split}}

\def\be{\begin{equation}}
\def\ee{\end{equation}}

\newcommand{\bea}{\begin{eqnarray}}
\newcommand{\eea}{\end{eqnarray}}

\journal{Physics Letters B}
\begin{document}
\begin{flushright}
{BRX-TH 6285}\\
CALT-TH 2014-156
\end{flushright}
\begin{frontmatter}
\title{Duality invariance of  $s\ge\frac{3}{2}$ fermions  in A\mbox{\rm d}S}
\author[Stanley]{S.~Deser}
\ead{deser@brandeis.edu}
\author[Florence]{D.~Seminara}
\ead{seminara@fi.infn.it}
\address[Stanley]{Walter Burke Institute for Theoretical Physics, California Institute of Technology, 
      Pasadena, CA 91125 and 
Physics Department, Brandeis University, Waltham, MA 02454, USA}
\address[Florence]{Dipartimento di Fisica, Universit\`a di Firenze and INFN Sezione di Firenze,
Via G. Sansone 1, 50019 Sesto Fiorentino, Italy}

\begin{abstract}
We  show that in $D=4$ AdS,  $\mbox{s}\ge 3/2$  partially massless (PM) 
fermions retain the duality invariances of their flat space massless counterparts. They have 
tuned  ratios  $ {m^2}/{M^2}\ne 0$
that turn them into sums of effectively massless 
unconstrained helicity $\pm(s,\cdots,\frac{3}{2})$ excitations, shorn of the lowest (non-dual)—helicity
$\pm \frac{1}{2}$-rung and -more generally- of succeeding higher rung as well. Each helicity mode is separately duality invariant, like its flat space
 counterpart.
\end{abstract}
\end{frontmatter}

\renewcommand{\thefootnote}{\arabic{footnote}}
\setcounter{footnote}{0}

\section{Introduction}

\noindent
 We address, and complete the answer to, the question whether/how free $m=0$ spin 
 $\ge 1$ systems can retain their known universal $(D=4)$ flat space duality 
 invariance \cite{Deser:2004xt} when embedded in (A)dS, 
 rather than flat, backgrounds. Half of the question had actually 
 already been answered in \cite{Deser:2013xb}, where it was shown that 
 the novel -- in dS -- PM irreps \cite{Deser:2001xr}  for $s>1$ bosons did so 
 (photons always do). The only difference from flat space was, perhaps surprisingly,
 that rather than having $m=0$ and just helicities ±s, they now sported a complete
  range, $\pm(s,\cdots,1)$ of effectively $m=0$ helicity excitations, excluding precisely 
  the helicity $0$ rung that would have spoiled the duality invariance manifest 
  in each of the higher ones. [Spin 1, being conformally invariant, is a 
  degenerate case, since (A)dS is conformally flat; of course, if studied exactly 
  like its $s>1$ peers in dS, its duality invariance follows exactly like theirs.
   Similarly, $s=3/2$ duality invariance was also exhibited long ago \cite{Deser:1977ur} for its
   ``massive'', cosmological SUGRA,  version.
  
For orientation, we recall that the easiest bosonic PM route of \cite{Deser:2013xb}
uses the dS frame $ds^2= -dt^2+ e^{2 M t} d\ell^2$, $M^2\equiv\frac{\Lambda}{3}$; 
there, one first discovers that, in "maximal" PM, a particular m/M ratio eliminates
 helicity-$0$, leaving a sum of unconstrained helicity $\pm (s.\cdots.1)$ actions. 
 Specifically, for the first non-trivial, $s=2$ model, the action is that of a 
 transverse-traceless (TT) spatial tensor and a transverse vector $(T_i)$; effective 
 masslessness is achieved by the PM tuning of the two mass parameters $(m,M)$.
  removing the helicity $0$ mode through the residual local scalar gauge invariance
  of the original action at the PM point, $m^2 = M^2$.
 However, as we shall see, $m=0$ models are 
 NOT duality invariant in (A)dS, because their lowest $(0 ~{\rm or}~ 1/2)$ helicities are reinstated there. 
 It was strongly 
 conjectured that the same process (also explicitly performed for $s=3$) goes 
 through for ALL $s$: 
the auxiliary fields, 
 constraints, etc that necessarily decorate the original covariant actions are 
 gone in the final, non-covariant, unconstrained 3+1 form.

Our spinor models also enjoy PM irreps, but in AdS instead of dS. As mentioned, 
$s=3/2$ is the basic, and long known, example of a dual invariant $s=3/2$ 
tuned system \cite{Deser:1977ur}: In order to obtain the cosmological, necessarily
 AdS— extension of SUGRA, one must add a mass term  $\sim m \bar\psi_m \sigma^{mn} 
 \psi_n$ to its massless action, with the tuning $m\sim 
 \sqrt{- \frac{\Lambda}{3}}=M$. This is exactly equivalent to improving the
 covariant derivative from $D_\mu$ to
 $D^\prime_\mu= D_\mu +\frac{m}{2} \gamma_\mu$. The effect of this change is 
 to restore the flat space commutativity, $[\partial_\mu, \partial_\nu]=0\to
  [D^\prime_\mu, D^\prime_\nu]=0$, thereby restoring the flat space invariance 
  of the model under local spinor transformations, now under $\delta \psi_\mu = 
  D^\prime_\mu \alpha(x)$, and  so again removing the lowest, helicity $1/2$, 
  excitation \cite{Deser:1976eh}. Here, the governing variables are the tranverse-traceless and 
  $\gamma_i-$traceless  spinor- spatial tensors $\psi^{tTT}_{ij\cdots}$, The PM 
  invariance \cite{Deser:2001xr} always removes the lowest, here helicity $1/2$, 
  leaving  
  an effectively massless (upon, legally, field redefining) array of helicities
   $\pm (s+1/2,…3/2)$, each separately duality invariant, but now at the 
above AdS point.  

\section{Derivation}
 For compactness, we will freely use the equations and results of 
 \cite{Deser:2001wx}; while that work is ostensibly formulated in dS,
 it is, as noted there, applicable to our AdS format, the  change in sign 
 of $\Lambda$ corresponds to setting the $M$ there to $i M$; we will simply 
 keep the $dS$ notation on the above understanding, rather than wasting 
 space with AdS formalism; we also borrow from \cite{Deser:2013xb} it 
 the near-certainty  
 that the procedure and results are uniform for all higher spins: again, 
 while higher spin actions require auxiliary fields and constraint variables,
  these are all  absent from the final unconstrained physical  3+1 actions, 
  here for the gamma-and spatial gradient-transverse, traceless spatial 
  tensor-spinor components.
  
   The key, Dirac, equation satisfied by these amplitudes
 is given by Eq. (14) there:  
 \be
 \label{(1)}
 \gamma^0\partial_0 \psi^{tTT}+e^{-M t} \cancel{\nabla} \psi^{tTT}+
 \left[m+(2-s)\gamma^0{M}\right]\psi^{tTT}=0
 \ee  
  Clearly, \eqref{(1)} differs from its flat space, massless, counterpart in two basic
  respects: it contains a ``mass'' term, $\sim a M\gamma^0+b~m $, as well as 
  the factor $e^{-M t}$ in the spatial derivative term. The latter is essentially 
  an irrelevant numerical coefficient in the Hamiltonian for spatial duality 
  transformation purposes, also  present and harmlessâ for bosons, as explained 
  in \cite{Deser:2013xb}. To remove the offending ``mass'' terms, 
consider for concreteness $s=5/2$\footnote{Different  values of the $s$ can be reduced to this case by performing the field the redefinition $\psi^{tTT}\mapsto e^{(s-5/2) M t}\psi^{tTT}$.  }, where we face
$-\frac{M}{2}\gamma^0 +m= \frac{M}{2} \gamma^0+ (m-\gamma^0 M)$. The parenthesis vanishes 
at the PM point, because $\gamma^0$ is diagonal with $\pm i$ entries, provided we add,
beyond $m^2+M^2=0$, the fermionic requirement that the upper/lower components 
of $\psi$ obey its respective roots $\pm i m+M=0$. 
 The remaining, 
$M/2\gamma^0$, term is simply removed by rescaling $\psi$ by $\exp(\frac{M}{2} t)$, to leave 
the sum of flat space pure helicity $>1/2$ actions (modulo the irrelevant $\mathrm{e}^{-M t}$ term 
in the Hamiltonian). While one might worry that any amount of M-dependence can be removed this way, the process here is really an artifact of the 
A{\rm d}S gauge choice: had we proceeded in conformal AdS gauge,
\be
 \label{(2)}
ds^2= (MT)^{-2} (-dT^2 + d\ell^2),
\ee
from the start, we would have found the fully ÒflatÓ form of (1),
\be
 \label{(3)}
\left[\gamma^0 \frac{\partial}{\partial T}+\cancel{\nabla}\right] \psi^{tTT}=0,
\ee
since PM actions are all conformally invariant (indeed, that is their special virtue). We can also recover \eqref{(3)} from the, final, massless \eqref{(1)}, 
 \be
  \label{(4)}
[\exp(Mt) \gamma^0 \partial_0 + \cancel{\nabla}] \psi^{tTT}=0, 
\ee
by  performing the (trivial) gauge transformation from our $t-$ to the $T-$frame \eqref{(2)}. But \eqref{(4)} is just the flat space,ÒE-BÓ, form given in \cite{Deser:1977ur} namely  
\be
\gamma^0 E + B=0.
\ee
This is both manifestly ($E\leftrightarrow B$) rotation invariant, and a time-local canonical transformation, in terms of the underlying canonical pair, as detailed in    \cite{Deser:2004xt}  for all spins. Indeed, the same, natural, conformal frame could have been used for the bosonic case [2] directly, or also reached by transforming to T-frame there, to remove the exp (MT) factor in the corresponding t-frame Hamiltonian there, starting from its PM form,
\be
{\cal L}_{\rm boson} =p^a \dot{q}_a - \exp (-MT) \frac{1}{2}[ p^2+q^2] 
\ee
where the summed index a runs over all helicities $>0$
(or $>1$ etc., in the various other PM levels discussed below).

We remark finally that for spins $>3/2$, there 
  exist different PM levels, each excising more lower helicities, until only
helicity $\pm s$ is left. Each of these occurs at different m/M ratios, and each is duality 
invariant by tuning $M$ to remove the 
$m$ term in the corresponding Dirac equation, then rescaling the spinor-tensor amplitude to 
remove whatever 
$M-$dependence remains. In this sense there is in fact a much larger set of dual-invariant 
PM$-$levels for 
any $s$, than 
the unique $m=0$ one in flat space. 

 \section{Summary}
 We have shown that all $s\ge 3/2$ PM free fermionic models in  suitably PM tunedâ  
 AdS are duality-invariant under the same transformations as in flat space,
  separately for each effectively massless helicity  $(>1/2)$ component.
  Together with the existing$-$essentially identical bosonic PM duality invariances in dS 
  \cite{Deser:2013xb}, this establishes the maximal curved spacetime generalization of flat 
  space free higher spin field duality invariances. 
\section*{Acknowledgements}
We happily thank A. Waldron for an encouraging discussion.  SD's work was supported in part by  NSF PHY- 1266107 and DOE \# DE-SC0011632 grants and DS's by  INFN, and COST Action MP1210 {\it ``The String Theory Universe"}.


\medskip
\begin{thebibliography}{99}
    \bibitem{Deser:2004xt} 
  S.~Deser and D.~Seminara,
  Phys.\ Lett.\ B {\bf 607}, 317 (2005)  [hep-th/0411169].
    \bibitem{Deser:2013xb} 
  S.~Deser and A.~Waldron,
  Phys.\ Rev.\ D {\bf 87}, 087702 (2013)
  [arXiv:1301.2238 [hep-th]].
   \bibitem{Deser:2001xr} 
  S.~Deser and A.~Waldron,
  Phys.\ Lett.\ B {\bf 513}, 137 (2001)
  [hep-th/0105181].
    \bibitem{Deser:1977ur} 
  S.~Deser, J.~H.~Kay and K.~S.~Stelle,
  Phys.\ Rev.\ D {\bf 16}, 2448 (1977).

    \bibitem{Deser:1976eh} 
  S.~Deser and B.~Zumino,
  Phys.\ Lett.\ B {\bf 62}, 335 (1976).
    \bibitem{Deser:2001wx} 
  S.~Deser and A.~Waldron,
  Phys.\ Lett.\ B {\bf 508}, 347 (2001)
  [hep-th/0103255].
\end{thebibliography}
\end{document}